\documentclass[12pt]{article}
\usepackage{amsmath,amscd,amsfonts,amssymb,latexsym}
\usepackage{epsfig}
\usepackage{graphicx}

\newcommand{\sect}[1]{\setcounter{equation}{0}
\section{#1}}

\def\ds{\displaystyle}
\def\be{\begin{equation}}
\def\ee{\end{equation}}

\begin{document}


\thispagestyle{empty}
\hfill \today

\vspace{2.5cm}

\begin{center}
\bf{\LARGE
Contractions from $osp(1|32) \oplus osp(1|32)$ to
\\[0.3cm] the M-theory superalgebra extended by
\\[0.3cm] additional fermionic generators}
\end{center}

\vspace{1cm}

\begin{center}
J.J. Fern\'andez, J.M. Izquierdo and M.A. del Olmo~$^1$
\end{center}

\begin{center}
{\sl Departamento de F\'{\i}sica Te\'orica  and $^1$~IMUVA, Universidad de
Valladolid, \\
E-47011, Valladolid, Spain.}\\
\medskip

{e-mail: julio.j.fernandez@hotmail.es, izquierd@fta.uva.es, olmo@fta.uva.es}

\end{center}

\vspace{1cm}

\begin{abstract}
We study here the generalized Weimar-Woods contractions of the superalgebra $osp(1|32) \oplus osp(1|32)$ in order to obtain a suitable algebra that could describe the gauge group of $D=11$ supergravity. The contracted superalgebras are assumed to be given in terms of fermionic extensions of the M-theory superalgebra. We show that the only superalgebra of this type obtained by contraction is the only one for which the three-form of $D=11$ supergravity cannot be trivialized. Therefore, $D=11$ supergravity cannot be connected in this way with a contraction of $osp(1|32) \oplus osp(1|32)$.
\end{abstract}
\vspace{1cm}

\newpage

\sect{Introduction}

 In their original work on $D=11$ supergravity, Cremmer, Julia and Scherck \cite{cremmerjulia} conjectured that the theory could admit a geometrical interpretation in terms of the simple supergroup $OSp(1|32)$. The evidence in favour of this conjecture was the fact that $OSp(1|32)$ contains an   $SO(8)$ subgroup, which is also a subgroup of the internal invariance group of a  $D=4$ reduction of the $D=11$ model.

However, the presence in the $D=11$ supergravity field content of a three-form field $A_{3}=A_{\mu\nu\rho}(x)\ dx^{\mu} \wedge dx^{\nu} \wedge dx^{\rho}$, apart from the elfbein $e^{a}=e^a_\mu dx^\mu$, the Rarita-Schwinger field  $\psi^{\alpha}= \psi^{\alpha}_\mu dx^\mu$ and the spin connection $w^{ab}= w^{ab}_\mu dx^\mu$ one-forms, makes  it difficult to identify the group theoretical structure of the model.  D'Auria and Fr\'e  \cite{dauriafre}, addressed this problem by looking at the free differential algebra (FDA) satisfied by the above forms in the absence of curvatures. This FDA does not  consist only of one-forms, so it cannot be interpreted as the Maurer-Cartan equations of a certain Lie superalgebra. D'Auria and Fr\'e's idea was to express the three-form $A_{3}$ in terms of linear combinations of exterior products of one-forms for a certain superalgebra, which had to be found. Two superalgebras were obtained which allowed the decomposition of $A_{3}$ in such a way. For this, it was necessary to consider superalgebras that, ignoring the $spin(10,1)$ algebra, are fermionic central extensions of  the M-theory superalgebra \cite{townsend,sezgin}, with anticommutators:
\begin{align}\label{equation1}
 \left\{Q_{\alpha},Q_{\beta}\right\}  =   \, P_{a}\ \Gamma^{a}_{\alpha \beta}\,
\, +\,\frac{1}{2!}\ Z_{ab}\ \Gamma^{ab}_{\alpha \beta}\,
\, +\ \frac{1}{5!}\ Z_{a_{1} \dots a_{5}}\ \Gamma^{a_{1} \dots a_{5}}_{\alpha \beta} \ .
\end{align}
An obvious question was  whether these two superalgebras could be related to $osp(1|32)$ or, more generally, to a simple superalgebra. This was considered in \cite{castellanifre}, where not only $osp(1|32)$, but also $osp(1|64)$ and $su(32|1)$, were ruled out as algebras that could lead to the D'Auria-Fr\'e ones by contraction.

The semisimple superalgebra $osp(1|32) \oplus osp(1|32)$ was not in the above list, but this is the algebra that was later considered by Horava \cite{horava} in he conjecture that M-Theory is a Chern-Simons (CS) field theory based on $osp(1|32) \oplus osp(1|32)$. An implication of Horava's conjecture is that $D=11$ Cremmer-Julia-Scherck supergravity would be a low-energy limit of a CS theory based on $osp(1|32) \oplus osp(1|32)$.

 In \cite{horava,nastase}, it was assumed that the supersymmetry group in the low-energy limit  had to be a contraction of $osp(1|32) \oplus osp(1|32)$. The contraction problem was considered in \cite{nastase}, where the superalgebras obtained, although with the same structure, did not coincide with those originally found by D'Auria and Fr\'e. The question remained of interpreting this discrepancy.

The two superalgebras found in \cite{dauriafre} were shown to be just two examples of an infinite set of them \cite{bandos04,bandos05}, which solved in general the problem posed by D'Auria and Fr\'e. In \cite{bandos04}, it was shown that all superalgebras with  structure
$\mathfrak{G}(s)={\mathfrak{E}}^{(528|32+32)}(s) \rtimes so(10,1)$,  where ${\mathfrak{E}}^{(528|32+32)}(s)$ is a fermionic central extension of the M-Theory superalgebra and $\rtimes$ indicates semidirect product of algebras, are actually parametrized by a real parameter $s$, and that for all values of $s$ except one, say $s = 0$ (see later), it was possible to interpret the three-form field $A_{3}$ in terms of one-forms dual to the generators of the algebra. This particular value of $s$, for which it is not possible to decompose  $A_{3}$, corresponds to the only superalgebra
 $\mathfrak{G}(0)={\mathfrak{E}}^{(528|32+32)}_{(s = 0)} \rtimes so(10,1)$ for which the Lorentz group $SO(10,1)$ can be enlarged to $Sp(32)$. Moreover, it is given by an expansion of $osp(1|32)$. Under another name, Lie algebra expansions were used for the first time by Hatsuda and Sakaguchi \cite{hatsuda} in order to relate the Wess-Zumino terms of the adS-type superstrings with the Poincar\'e ones. The expansion method, studied in full generality in \cite{azcarraga03}, consists of expan\-ding the Maurer-Cartan dual one-forms of an initial superalgebra in terms of a parameter $\lambda$, and then identifying the coefficients of each power in $\lambda$ in the resulting Maurer-Cartan equations\footnote{See \cite{izaurieta}, for  generalization of the expansion approach introducing semigroups}. In this way, superalgebras with an infinite number of generators are obtained. Certain conditions can be imposed in order to ensure that, by cutting the expansion in $\lambda$ up to a finite power, the resulting equations are the MC equations of a finite (super)algebra  \cite{azcarraga03}. It turns out that ${\mathfrak{E}}^{(528|32+32)}_{(s = 0)} \rtimes so(10,1)$ corresponds to the expanded superalgebra $osp(1|32)(2,3,2)$ (see \cite{bandos04,bandos05,azcarraga03} for the notation).

In all the examples obtained so far, the resulting expansions can be viewed as extensions followed by contractions, and this will presumably be true in general. The inverse statement is obviously false, because an arbitrary extension plus a contraction does not have to be an expansion since expansions remember the structure of the original, unexpanded algebra. So, it makes sense to find out whether the contraction of $osp(1|32) \oplus osp(1|32)$ leads to an expansion of $osp(1|32)$ or to the other ($s\neq 0$)  superalgebras in the class of fermionic extensions of the M-theory superalgebras, $\mathfrak{G}(s\neq 0)$. We have performed a systematic calculation of all possible contractions of $osp(1|32) \oplus osp(1|32)$ leading to a superalgebra with the genertic structure
\begin{align}\label{equation2}
\left({\mathfrak{E}}^{(528|32+32)}(s) \oplus \mathcal{L}^{(473)}\right)  \rtimes so(10,1)\ ,
\end{align}
where $\mathcal{L}$ is an arbitrary superalgebra that has to be present because the contraction procedure does not change the dimension of the  superalgebra and the dimensions of $osp(1|32) \oplus osp(1|32)$ and those of the $\mathfrak{G}(s)$ do not match. Indeed, $osp(1|32) \oplus osp(1|32)$ has dimension $1120=2\times\left(\frac{32\times 33}{2} +32 \right) = 2\times \left( 11+ {11\choose 2}+ {11\choose 5} +32 \right)$, whereas $ {\mathfrak{E}}^{(528|32+32)}(s)  \rtimes so(10,1)$ have dimension $647= 528+ 32 +32 + {11\choose 2}$, so the dimension of the bosonic Lie algebra $\mathcal{L}$ is equal to $473$.

The main result of this paper is that it is possible to obtain by contraction only the fermionic extension of the M-theory superalgebra given by an expansion, {\it i.e.} the case  $s = 0$  in \eqref{equation2}. In other words, none of the Lie superalgebras, suitable for decomposing the three-form $A_{3}$ of $D=11$ supergravity in terms of MC one-forms, can be obtained by contraction of the mixture of two $osp(1|32)$ algebras.

The plan of the paper is as follows: in section~\ref{superalgebraosp132} a review of $osp(1|32)$ and the fermionic extensions of the M-theory algebras is presented. Section~\ref{contractions} explains the procedure used to obtain the contractions of $osp(1|32) \oplus osp(1|32)$  and contains the statement of the main result of this paper. Finally, section~\ref{conclusions} is devoted to the conclusions.

\sect{The superalgebra $osp(1|32)$}\label{superalgebraosp132}

The orthosymplectic Lie algebra $osp(1|32)$ can be defined, in a certain basis $\left\{Z_{\alpha \beta},\ Q_{\gamma}\right\}$, by the following anticommutators and commutators relations:
\begin{equation}\label{equation3}
\begin{array}{lll}
\left\{Q_{\alpha},Q_{\beta}\right\} &=& \eta \ Z_{\alpha\beta}\ ,  \\[0.3cm]
[Z_{\alpha \beta},Q_{\gamma}] &=& C_{\alpha \gamma}\ Q_{\beta}\ +\ C_{\beta \gamma}\ Q_{\alpha}\ , \\[0.3cm]
 [Z_{\alpha\beta}, Z_{\gamma\delta}] &=& C_{\alpha\gamma} Z_{\beta\delta} +
 C_{\beta\gamma} Z_{\alpha\delta}
+  C_{\alpha\delta} Z_{\beta\gamma} +  C_{\beta\delta} Z_{\alpha\gamma}\ ,
\end{array}\end{equation}
where $Z_{\alpha\beta}$ is a symmetric matrix in the spinorial indices $(\alpha, \beta, \gamma  = 1, \dots ,32)$, $C_{\alpha \beta}$ is the $32\times32$ skewsymmetric charge conjugation matrix  and $\eta=\pm1$. Notice that both values of $\eta$  do not make any difference in the complex Lie algebra, but in the real case  they determine  non isomorphic superalgebras denoted by $osp_{+}(1|32)$ and $osp_{-}(1|32)$, as is the case for $osp(1|2) (see $\cite{achucarro}).
The above (anti-)commutators are dual to the following Maurer-Cartan equations (see, {\it e.g.} \cite{14})
\begin{equation}\label{equation4}
\begin{array}{lll}
d\Pi_{\alpha \beta} &=& -\  (\Pi_{\alpha \gamma} \wedge \Pi^{\gamma}\ _{\beta}) \ - \eta \  (\Pi_{\alpha} \wedge \Pi_{\beta}) \ , \\[0.3cm]
d\Pi_{\alpha}&=& -\  \Pi_{\alpha \gamma} \wedge \Pi^{\gamma}\ ,
\end{array}\end{equation}
 where $\Pi_{\alpha \beta}$ ($\Pi_{\alpha}$) are the MC one-forms dual to $Z_{\alpha \beta}$ ($Q_{\alpha}$)
 \[
 \Pi^{\alpha \beta}(Z_{\gamma\delta})=2\delta^{(\alpha}_\gamma
\delta^{\beta)}_{\delta} \equiv \delta^{\alpha}_\gamma
\delta^{\beta}_{\delta} + \delta^{\beta}_\gamma
\delta^{\alpha}_{\delta},\quad\qquad  \Pi^\alpha(Q_\beta)= \delta^\alpha_\beta.
\]

Since we are interested in contractions that treat differently the various $SO(1,10)$ Lorentz components of $Z_{\alpha \beta}$ (or $\Pi_{\alpha \beta}$), we
 express  $Z_{\alpha\beta}$ in terms of the tensorial generators $Z_{a},\ Z_{ab},\ Z_{a_{1}\dots a_{5}}$ (or $\Pi_{a},\ \Pi_{ab},\ \Pi_{a_{1}\dots a_{5}}$), with $a=0,\dots,10$, by using the basis of the  11-dimensional Dirac matrices $\Gamma^{a}_{\alpha \beta},\ \Gamma^{ab}_{\alpha \beta},\ \Gamma^{a_{1}\dots a_{5}}_{\alpha \beta}$, as
\begin{align}
\label{equation5}
Z_{\alpha \beta} = \frac{1}{1!\cdot 32}\ \Gamma^{a}_{\alpha \beta}\ Z_{a} + \frac{1}{2!\cdot 32}\ \Gamma^{ab}_{\alpha \beta}\ Z_{ab} + \frac{1}{5!\cdot32}\ \Gamma^{a_{1}\dots a_{5}}_{\alpha \beta}\ Z_{a_{1}\dots a_{5}}\ ,
\end{align}
and similarly for $\Pi_{\alpha \beta}$, where  the notation $\Gamma^{a_{1}\dots a_{n}}_{\alpha \beta}$ refers to $(\Gamma^{a_{1}\dots a_{n}} C^{-1})_{\alpha \beta}$ with \ $C^{-1}$ being  the inverse of the charge conjugation matrix .

Using the relation (\ref{equation5}) in (\ref{equation3}),  we obtain the commutators and anti-commutators relations \cite{15,16}
\be\label{equation6}
\begin{array}{lll}
[Z_{a},Z_{b}]& =& \ds\frac{1}{8}\ J_{ab}\ ,  \\[0.3cm]
[Z^{a},J_{b_{1}b_{2}}] &=&\ds \frac{1}{4}\ \delta_{[b_{1}}^{a}\;
\delta_{b_{2}]}^{k}\ Z_{k}\ ,  \\[0.4cm]
[J^{a_{1}a_{2}},J_{b_{1}b_{2}}] &=& \ds\frac{1}{2}\ \delta_{[k_{1}}^{[a_{1}}\ \delta_{[b_{1}}^{a_{2}]}\; \delta_{b_{2}]}^{k_{2}]}\ J^{k_{1}}\ _{k_{2}}\ , \\[0.4cm]
[Z_{a},Z_{b_{1} \dots b_{5}}] &=& \ds\frac{i}{8 \cdot5!}\ \epsilon_{c_{5} \dots c_{1} k_{1} {k_{2} \dots k_{6}}}\;\delta_{a}^{[k_{1}}\delta_{[b_{1}}^{k_{2}} \dots \delta_{b_{5}]}^{k_{6}]}\ Z^{c_{1} \dots c_{5}}\ , \\[0.4cm]
[J^{a_{1} a_{2}},Z_{b_{1}\dots b_{5}}] &=&\ds \frac{5}{4}\ \delta_{[k_{1}}^{[a_{1}}\ \delta_{[b_{1}}^{a_{2]}}\;\delta_{b_{2}}^{k_{2}}\dots \delta_{b_{5}]}^{k_{5}]}\ Z^{k_{1}}\  _{k_{2}\dots k_{5}}\ ,\\[0.4cm]
[Z^{a_{1} \dots a_{5}},Z_{b_{1} \dots b_{5}}] &=& \ds\frac{i}{8}\ \delta_{[k_{1}}^{[a_{1}} \dots \delta_{k_{5}}^{a_{5}]} \delta_{[b_{1}}^{k_{6}} \dots \delta_{b_{5}]}^{k_{10}]}\ \epsilon^{k_{1} \dots k_{5}}\ _{k_{6} \dots k_{10}c}\; Z^{c}
\\[0.3cm]
&&\quad \ds+\; \frac{5i}{4!}\ \delta_{[k_{1}}^{[a_{1}} \delta_{k_{2}}^{a_{2}} \delta_{k_{3}}^{a_{3}} \delta_{[b_{2}}^{a_{4}}\delta_{b_{1}}^{a_{5}]}
\delta_{b_{3}}^{[k_{4}}\delta_{b_{4}}^{k_{5}} \delta_{b_{5}]}^{k_{6}]}
\epsilon^{k_{1}k_{2}k_{3}}\ _{k_{4}k_{5}k_{6}c_{5}c_{4}c_{3}c_{2}c_{1}}\\[0.3cm]
&&\qquad\ds \times Z^{c_{1}c_{2}c_{3}c_{4}c_{5}}
 +\;75\ \delta_{[k_{1}}^{[a_{1}}\delta_{[b_{4}}^{a_{2}} \dots \delta_{b_{1}}^{a_{5}]} \delta_{b_{5}]}^{k_{2}]} J^{k_{1}}\ _{k_{2}}\ ,\\[0.4cm]
[Z_{a},Q_{\alpha}] &=&\ds \frac{1}{16}\ (\Gamma_{a})_{\alpha}\ ^{\beta}\ Q_{\beta}\ , \\[0.4cm]
[J_{ab},Q_{\alpha}] &=&\ds -\frac{1}{16}\ (\Gamma_{a b})_{\alpha}\ ^{\beta}\ Q_{\beta}\ ,   \\[0.4cm]
[Z_{a_{1} \dots a_{5}},Q_{\alpha}] &=&\ds \frac{1}{16}\ (\Gamma_{a_{1} \dots a_{5}})_{\alpha}\ ^{\beta}\ Q_{\beta}\ ,
 \\[0.4cm]
\left\{Q_{\alpha},Q_{\beta}\right\} &=&\ds \Gamma^{a}_{\alpha \beta}\ Z_{a}
\ +\ \frac{1}{2!}\ \Gamma^{ab}_{\alpha \beta}\ J_{ab}
\ +\ \frac{1}{5!}\ \Gamma^{a_{1} \dots a_{5}}_{\alpha \beta}\ Z_{a_{1} \dots a_{5}}\ ,
\end{array}
\ee
 where the antisymmetrisation in the r.h.s., denoted by the square brackets, is such that the overall weight is 1. Correspondingly,  the analogue to equation (\ref{equation5}) for $\Pi_{\alpha \beta}$ in (\ref{equation4}) is given by
the Maurer-Cartan equations
\be\label{equation7}
\begin{array}{lll}
d\Pi^{a}&=&\ds -\frac{1}{8}\ (\Pi^{b} \wedge \Pi_{b}\ ^{a}) - \frac{1}{2}\ \Gamma^{a}_{\alpha \beta}\ (\pi^{\alpha}\wedge \pi^{\beta})  \\[0.3cm]
&&\qquad \ds- \frac{i}{16 \cdot (5!)^2} \ \epsilon^{a\, b_{1} \dots b_{5}}\ _{c_{1} \dots c_{5}}\ (\Pi_{b_{1} \dots b_{5}}\wedge \Pi^{c_{1} \dots c_{5}})\ ,\\[0.4cm]
d\Pi^{ab}&=&\ds -\frac{1}{8}\ (\Pi^{a}\wedge \Pi^{b}) - \frac{1}{8}\ (\Pi^{ac}\wedge \Pi_{c}\ ^{b})   \\[0.3cm]
&&\qquad \ds- \frac{1}{2} \Gamma^{ab}_{\alpha \beta}\ (\pi^{\alpha}\wedge \pi^{\beta})
- \frac{1}{4!\cdot 8}
\; (\Pi^{a}\ _{c_{1} \dots c_{4}}\wedge \Pi^{c_{1} \dots c_{4}\, b})\ ,
\\[0.4cm]
d\pi^{\alpha} &=&\ds \frac{1}{16}\ (\Gamma_{a})_{\beta}\ ^{\alpha}\ (\pi^{\beta}\wedge \Pi^{a}) - \frac{1}{2\cdot16}\ (\Gamma_{ab})_{\beta}\ ^{\alpha}\ (\pi^{\beta}\wedge \Pi^{ab})  \\[0.3cm]
&&\qquad\ds+ \frac{1}{5!\cdot 16} (\Gamma_{a_{1}\dots a_{5}})_{\beta}\ ^{\alpha}\ (\pi^{\beta}\wedge \Pi^{a_{1}\dots a_{5}})\ ,  \\[0.4cm]
d\Pi^{a_{1}\dots a_{5}} &=&\ds -\frac{i}{5!\cdot 8}\ \epsilon_{c\, b_{1}\dots b_{5}}\ ^{a_{1}\dots a_{5}}\ (\Pi^{c}\wedge \Pi^{b_{1}\dots b_{5}})\ - \frac{5}{8}\ (\Pi^{[a_{1}}\ _{b} \wedge \Pi^{b\, a_{2}\dots a_{5}]})  \\[0.3cm]
&&\qquad\ds-\frac{1}{2}\ \Gamma^{a_{1}\dots a_{5}}_{\alpha \beta}\ (\pi^{\alpha}\wedge \pi^{\beta})\\[0.3cm]
&&\qquad\qquad\ds
-\frac{i}{2\cdot(4!)^2}\ \epsilon^{a_{1}\dots a_{5} b_{1}b_{2}b_{3}}\ _{c_{1}c_{2}c_{3}}\ (\Pi_{b_{1}\dots b_{5}}\wedge \Pi^{b_{5}b_{4}c_{1}c_{2}c_{3}})\ .
\end{array}\ee
We will use, in what follows, eqs. (\ref{equation7}) rather than the equivalent Lie superalgebra commutators of eqs. (\ref{equation6}).
The algebras that we are going to obtain are related with $\mathfrak{G}(s)={\mathfrak{E}}^{(528|32+32)}_{(s)} \rtimes so(10,1)$ \cite{bandos04,bandos05}. They are determined by the generators $Q_{\alpha},\ Q'_{\alpha},\ Z_{a},\ Z_{ab},\ Z_{a_{1} \dots a_{5}}$, plus the Lorentz generators $J_{ab}$, and their commutators and anti-commutators can be given in the form
\be\label{equation8}
\begin{array}{lll}
[Z_{a},Q_{\alpha}]  &=&\ds \tau_{2}\ (s-1)\  (\Gamma_{a})_{\alpha}\ ^{\beta}\ Q^{'}_{\beta}\ , \\[0.4cm]
 [Z_{ab} , Q_{\alpha}] &=&\ds \tau_{2}\ (\Gamma_{ab})_{\alpha}\ ^{\beta}\ Q^{'}_{\beta}\ , \\[0.4cm]
[Z_{a_{1} \dots a_{5}} , Q_{\alpha}]  &=&\ds \tau_{2}\ (\frac{s}{6!} - \frac{1}{5!})\ (\Gamma_{a_{1} \dots a_{5}})_{\alpha}\ ^{\beta}\ Q^{'}_{\beta}\ , \\[0.4cm]
\left\{Q_{\alpha},Q_{\beta}\right\}  &=&\ds \Gamma^{a}_{\alpha\beta}\ Z_{a} + \frac{1}{2!}\ \Gamma^{ab}_{\alpha\beta}\ Z_{ab} + \frac{1}{5!}\ \Gamma^{a_{1} \dots a_{5}}_{\alpha\beta}\ Z_{a_{1}\dots a_{5}}\ ,
\end{array}\ee
 plus the obvious ones involving the Lorentz generators given the Lorentz character of the algebra. Note that in eqs. (\ref{equation8}) the real parameter $\tau_{2}$ is always different from zero and it can be absorbed in the definition of $Q'_{\beta}$, so that only one free real  parameter $s$ remains. This factorization of the algebra also includes the case when $\tau_{2}\rightarrow 0$ and so $s\rightarrow \infty$, such that the product $\tau_{2}\cdot s$ remains finite.

The corresponding Maurer-Cartan equations are
\be\label{equation9}\begin{array}{lll}
d\Pi^{a}  &=&\ds -\frac{1}{2}\  \Gamma _{\alpha \beta}^{a}\ (\pi^{\alpha}\  \wedge \pi^{\beta})\ , \\[0.4cm]
d{\Pi'}\, ^{ab}  &=&\ds - \frac{1}{2}\ \Gamma _{\alpha \beta}^{ab}\ (\pi^{\alpha} \wedge \pi^{\beta})\ , \\[0.4cm]
d\Pi^{a_{1}\dots a_{5}}  &=&\ds -\ \frac{1}{2}\ \Gamma_{\alpha \beta}^{a_{1} \dots a_{5}}\ (\pi^{\alpha} \wedge \pi^{\beta})\ , \\[0.4cm]
d\pi^{\alpha}  &=&\ds 0\ , \\[0.4cm]
d{\pi'}\, ^{\alpha}  &=&\ds - \tau_{2}  \left(  (s-1) \ (\Gamma_{a})_{\beta}\ ^{\alpha} \ (\Pi^{a} \wedge \pi^{\beta}) + \frac{1}{2!} (\Gamma_{ab})_{\beta}\ ^{\alpha}\ (\Pi'\,^{ab} \wedge \pi^{\beta})\right. \\[0.3cm]
 &&\qquad\qquad\ds\left.+\; \left(\frac{s}{6!} - \frac{1}{5!}\right) \left(\Gamma_{a_{1} \dots a_{5}})_{\beta}\ ^{\alpha} (\Pi^{a_{1} \dots a_{5}} \wedge \pi^{\beta}\right)\right)\ ,
\end{array}\ee
where $\Pi^{a},\ \Pi'\,^{ab},\ \Pi^{a_{1}\dots a_{5}},\ \pi^{\alpha},\ \pi'\,^{\alpha}$ are the dual one-forms to the algebraic generators $Z_{a},\ Z_{ab},\ Z_{a_{1} \dots a_{5}},\ Q_{\alpha},\ Q'_{\alpha}$, respectively.

In this parametrization, all the algebras in (\ref{equation9}) (resp. \ref{equation8})  can be used to write the three-form $A_{3}$ of $D=11$ supergravity as a composite one, except for the case $s=0$, which coincides with a Lie algebra expansion of $osp(1|32)$ \cite{bandos04,bandos05,azcarraga03}.

\sect{Contractions of $osp(1|32) \oplus osp(1|32)$}\label{contractions}

Generalized, or Weimar-Woods \cite{22,23}, contractions can be constructed as follows: let $\mathcal{G}$ be a Lie (super)algebra
given, as a vector space, by the direct sum
\begin{equation}
\label{splitting}
\mathcal{G} = V_0 \oplus V_1 \oplus \dots \oplus V_n \ ,
\end{equation}
and such that the (graded) commutators obey
\begin{equation}
\label{gradcom}
[ V_p, V_q ] \subset \bigoplus^{p+q}_{l=0} V_l \ .
\end{equation}
In particular, $V_0$ is a subalgebra of $\mathcal{G}$. Let $^\{X_{p,\, \alpha_p}\}$, $p=0,\dots , n$, $\alpha_p=1,\dots , \dim V_p$,
be a basis of $\mathcal{G}$ relative to the splitting \eqref{splitting}. Then, the structure constants
$C^{r,\, \gamma_r}_{p,\, \alpha_p\; q,\, \beta_q}$ vanish for $r>p+q$. If $\omega^{p,\, \alpha_p}$ are the one-forms dual to the
vector fields $X_{p,\, \alpha_p}$, $\omega^{p,\, \alpha_p}(X_{q,\, \beta_q}) = \delta^p_q \delta^{\alpha_p}_{\beta_q}$, the
MC equations of $\mathcal{G}$ are then
\begin{equation}
\label{MCinitial}
d\omega^{r,\, \gamma_r} = - \frac{1}{2} \sum_{p+q\leq r} C^{r,\, \gamma_r}_{p,\, \alpha_p\; q,\, \beta_q} \omega^{p,\, \alpha_p} \wedge
\omega^{q,\, \beta_q}\ .
\end{equation}
It turns out that the same vector space \eqref{splitting}, but now with modified MC equations given by \eqref{MCinitial} with the sum only
extended to $p+q=r$, defines a new Lie (super)algebra $\mathcal{G}_c$, known as the Weimar Woods contracted /super)algebra
relative to the splitting \eqref{splitting}. This contracted algebra can be obtained by re-scaling in terms of a
parameter $\lambda$ the forms $\omega^{p,\, \alpha_p}$
as $\omega^{p,\, \alpha_p} \rightarrow \lambda^p \omega^{p,\, \alpha_p}$ in the starting MC equations, and then taking the
limit $\lambda \rightarrow 0$. This is the procedure that we use in this paper. The case $n=1$ corresponds to the
original, \.{I}n\"{o}n\"{u}-Wigner \cite{17,18}, contractions.

Contractions are dimension preserving. What it is not always realized is that the contraction of a direct sum of two Lie (super)-algebras $\mathcal{G} \oplus \mathcal{\bar{G}}$ can be different from the direct sum of the contractions of $\mathcal{G}$ and $\mathcal{\bar{G}}$, i.e.,
\begin{equation}
\label{equation10}
	(\mathcal{G} \oplus \mathcal{\bar{G}})_{c} \neq \mathcal{G}_{c} \oplus \mathcal{\bar{G}}_{c}\ .
\end{equation}
For instance, the non-trivially extended Galilei algebra may be obtained as a contraction of the trivial extension of the Poincar\'e algebra by $u(1)$ \cite{20,19}, and the superalgebra of $D=3$ $(p,q)$-Poincar\'e supergravity as a contraction of $osp_{+}(p|2) \oplus osp_{-}(q|2) \oplus so(p) \oplus so(q)$ \cite{21}. This situation happens when the contraction is performed relative to a basis that is a linear combination of generators in $\mathcal{G}$ and $\mathcal{\bar{G}}$, and the inverse of this linear combination is not defined in the contraction limit, so it cannot be undone after the contraction.

In our case we have that  $\mathcal{G} = osp_{+}(1|32)$ and $\mathcal{\bar{G}} = osp_{-}(1|32)$. Let $\{X_{i}\}$ and $\left\{\bar{X}_{i}\right\}$ be bases of generators of the Lie algebras $\mathcal{G}$ and $\mathcal{\bar{G}}$, respectively. We will considerer a new basis $\{Y_{i} , \bar{Y}_{i}\}$ of $\mathcal{G} \oplus \mathcal{\bar{G}}$ by
\be\label{equation11}
\begin{array}{lll}
Y_{i} &=&\ds A^{j}_{i} \ X_{j} + B^{j}_{i}\  \bar{X}_{j}\ , \\[0.4cm]
\bar{Y}_{i} &=&\ds C^{j}_{i} \ X_{j} + D^{j}_{i}\  \bar{X}_{j}\ ,
\end{array}\ee
and then perform a generalized Weimar-Woods contraction \cite{22,23} by rescaling in a consistent way the set of generators $\{Y_{i},\bar{Y}_{i}\}$ so that the contraction limit (when the scaling parameter goes to zero) is well defined. Alternatively, this can also be done with the Maurer-Cartan forms dual to the generators. In fact, this is how we have done our calculations. Since $osp_{+}(1|32)$ and $osp_{-}(1|32)$ are actually two non-isomorphic real versions of the same complex algebra, we can take $\mathcal{G} =\mathcal{\bar{G}} = osp_{+}(1|32)$ by considering complex coefficients $\left\{A^{j}_{i},B^{j}_{i},C^{j}_{i},D^{j}_{i}\right\}$. So, we shall take two copies of (\ref{equation5}), (\ref{equation6}) and (\ref{equation7}) (or (\ref{equation8}), (\ref{equation9})) and consider complex linear combinations of the generators, and then look for contractions that would correspond to a real superalgebra. The linear combinations will not be arbitrary, because the Lorentz character of the components of $Z_{\alpha\beta}$ in (\ref{equation10}) has to be preserved. This means that we will take combinations of $Z_{a}$ and $Z'_{a}$, $Z_{ab}$ and $Z'_{ab}$, etc. separately, with scalar coefficients. In terms of the Maurer-Cartan one-forms, we write generically
\be\label{equation12}
\begin{array}{lll}
\rho_{+}^{(n)} &=&\ds  \alpha_{(n)}\ \Pi^{(n)}+ \ \beta_{(n)}\ \bar{\Pi}^{(n)}\ ,\\[0.4cm]
\rho_{-}^{(n)} &=&\ds  \gamma_{(n)}\ \Pi^{(n)}+ \ \delta_{(n)}\ \bar{\Pi}^{(n)}\ ,
\end{array}\ee
where $n = (1,2,5,\alpha)$ denotes the number of  Lorentz indices  for the bosonic one-forms $\rho_{\pm}^{a}, \rho_{\pm}^{ab}, \rho_{\pm}^{a_{1}\dots a_{5}}$ or the spinorial index for the fermionic one $\rho_{\pm}^{\alpha} \equiv \psi_{\pm}^{\alpha}$.
We must  ensure that the linear combinations have to be invertible, so that we really perform a change of basis. Hence
\begin{equation}
\label{equation13}
det \begin{pmatrix}
\alpha_{(n)} & \beta_{(n)} \\
\gamma_{(n)} & \delta_{(n)}
\end{pmatrix} \neq 0
\end{equation}
Then, we pose the  problem of  finding the exponents of $\lambda$ used to do the following rescaling
\begin{equation}\label{equation14}
\begin{array}{llllllllllll}
\rho_{+}^{a}  &\Rightarrow& \lambda^{n}\; \rho_{+}^{a}, \;\; &
\rho_{+}^{ab}  &\Rightarrow& \lambda^{p}\;\rho_{+}^{ab}, \;\; &
\rho_{+}^{a_{1}\dots a_{5}}  &\Rightarrow& \lambda^{r}\ \rho_{+}^{a_{1}\dots a_{5}},\;\; &
\psi^{\alpha}_{+} &\Rightarrow& \lambda^{v}\ \psi^{\alpha}_{+},\\[0.4cm]
\rho_{-}^{a}  &\Rightarrow& \lambda^{m}\ \rho_{-}^{a},&
 \rho_{-}^{ab}  &\Rightarrow& \lambda^{q}\ \rho_{-}^{ab},&
\rho_{-}^{a_{1}\dots a_{5}}  &\Rightarrow& \lambda^{t}\ \rho_{-}^{a_{1}\dots a_{5}},&
\psi^{\alpha}_{-} &\Rightarrow& \lambda^{w}\ \psi^{\alpha}_{-},
\end{array}\end{equation}
and the coefficients of (\ref{equation12}) with the condition (\ref{equation13}), such that the generalized Weimar-Woods contraction limit of the Maurer-Cartan equations leads to the structure (\ref{equation2}).

We have performed the calculations by using a symbolic manipulation programme (Mathematica). The problem to solve becomes an algebraic system in the complex coefficients $\left\{\alpha_{(n)}; \beta_{(n)}; \gamma_{(n)}; \delta_{(n)}\right\}$, those of the matrices inverse to the matrix in \eqref{equation13}
 (see Appendix) and in the real parameter $s$. More explicitly,
after the redefinition \eqref{equation12} and the change of scale \eqref{equation14}, the r.h.s of the resulting Maurer-Cartan equations will be given by sums of terms with the following structure
\[
\label{structure}
      \lambda^{E(m,n,p,q,r,t,u,w)}\ C(\alpha_{(n)}, \beta_{(n)}, \gamma_{(n)}, \delta_{(n)})\
 (\rho_{\pm} \wedge \rho_{\pm}),
\]
i.e., apart from the exterior product of two one-forms, there is a power of  $\lambda$ that
depends on the scaling factors of \eqref{equation14}, and a coefficient (structure constant) that depends on the parameters of the linear combination \eqref{equation12}. We have first imposed the condition that some of these terms  reproduce the ones in \eqref{equation9}, which fixes the values of their $C's$
in terms of the real parameter $s$ and also implies that their corresponding exponents $E$ vanish.  As a result, some of the remaining exponents are negative so their coefficients $C$ have to vanish consistently with the Weimar-Woods approach. Additionally, there are also terms that should not appear in the limit $\lambda\rightarrow 0$, which means that for them either $E>0$ or $C=0$. Finally, as already mentioned, the linear combinations in \eqref{equation12} have to be invertible.

This process leads to a system of equations and inequations that turns out to have a solution only when $s=0$. We remind that this value corresponds in our parametrization to the expansion of $osp(1|32)$
for which the three-form of $D=11$ supergravity $A_{3}$ cannot be written in terms of Maurer-Cartan one-forms. We have also considered the case $s\rightarrow \infty$ and checked that there is no solution. We do not include here the detailed expressions of the explicit computing, but they are available from the authors upon request.

\sect{Conclusions}\label{conclusions}

We have shown in this paper that is not possible to obtain by generalized Weimar-Woods contraction from $osp_{+}(1|32) \oplus\ osp_{-}(1|32)$ none of the algebras found in \cite{bandos04}, which allow   a gauge group interpretation of the three-form field in the sense of \cite{dauriafre}. Hence, we can conclude that $D=11$ supergravity cannot be connected with the semi-simple supergroup $OSp_{+}(1|32) \otimes\ OSp_{-}(1|32)$ by trivializing the three-form field $A_{3}$.

This result, however, does not necessarily mean that the conjecture made in \cite{horava}, according to which $D=11$ supergravity can be obtained as a low-energy limit of a Chern-Simons theory based on $osp_{+}(1|32) \oplus osp_{-}(1|32)$, is incorrect. This is so because although the low-energy limit corresponds to a rescaling of the gauge fields in terms of a parameter $\lambda$ of the type that appears when a Weimar-Woods contraction is performed, only the leading or next to leading terms in the resulting expansion of the Chern-Simons action in powers of $\lambda$ have the symmetries of the contracted algebra. However, the term that would correspond to supergravity is neither the leading nor the next to leading term, hence it is unclear how the problem of checking the connection of Chern-Simons supergravity with the ordinary one would be related to the algebra contractions.

\section*{Appendix}

In this appendix we show the Maurer-Cartan equations of the  $osp_{+}(1|32) \oplus osp_{-}(1|32)$ superalgebra. These relations are written in terms of the complex scalar coefficients (\ref{equation12}) used to determine the linear combinations between the rescaled one-forms of the two $osp_{+}(1|32)$ algebras
\[\begin{array}{l}
\left\{\alpha_{(n)};\beta_{(n)};\gamma_{(n)};\delta_{(n)} \right\} \equiv
\begin{pmatrix}
\alpha_{(n)} & \beta_{(n)}\\
\gamma_{(n)} & \delta_{(n)}
\end{pmatrix}\\[0.6cm]
 \qquad\equiv
\left\{
\begin{pmatrix}
A & B\\
C & D\\
\end{pmatrix}_{n=1}
;\quad
\begin{pmatrix}
E & F\\
G & H\\
\end{pmatrix}_{n=2}
;\quad
\begin{pmatrix}
I & J\\
K & L\\
\end{pmatrix}_{n=5}
;\quad
\begin{pmatrix}
M & N\\
O & P\\
\end{pmatrix}_{n=\alpha}
\right\}
\end{array}\]
their inverse relations
\[\begin{array}{l}
\left\{\alpha'_{(n)};\beta'_{(n)};\gamma'_{(n)};\delta'_{(n)}\right\} \equiv
\begin{pmatrix}
\alpha'_{(n)} & \beta'_{(n)}\\
\gamma'_{(n)} & \delta'_{(n)}
\end{pmatrix}
\\[0.6cm]\qquad
 \equiv
\left\{
\begin{pmatrix}
a' & b'\\
c' & d'\\
\end{pmatrix}_{n=1}
;\quad
\begin{pmatrix}
e' & f'\\
g' & h'\\
\end{pmatrix}_{n=2}
;\quad
\begin{pmatrix}
i' & j'\\
k' & l'\\
\end{pmatrix}_{n=5}
;\quad
\begin{pmatrix}
m' & n'\\
o' & p'\\
\end{pmatrix}_{n=\alpha}
\right\},
\end{array}\]

 and the structure constants (\ref{equation7}).
These explicit relations are
\medskip

{\bf 1).- For one-index tensors}
\[\begin{array}{lll}
d{\rho_{+}^{a}} &=&\ds - \frac{1}{8} \left\{ \left(A\ a'\ e' + B\ c'\ g' \right) \left(\rho_{+}\thinspace^{b} \wedge \rho_{+}\thinspace_{b}\thinspace^{a}\right) + \left(A\, a'\, f' + B\ c'\ h' \right) \left(\rho_{+}\thinspace^{b} \wedge \rho_{-}\thinspace_{b}\thinspace^{a}\right)\right. \\[0.3cm]
&&\ds\;\;\qquad +\left. \left(A\, b'\, e' + B\ d'\, g' \right) \left(\rho_{-}\thinspace^{b} \wedge \rho_{+}\thinspace_{b}\thinspace^{a}\right) + \left(A\, b'\, f' + B\, d'\, h' \right) \left(\rho_{-}\thinspace^{b} \wedge \rho_{-}\thinspace_{b}\thinspace^{a}\right)\right\} \\[0.35cm]
&&\ds- \frac{1}{2} \left\{
 \left(A\, m'\, m' + B\, o'\, o'\right) \left(\psi_{+}\,^{\alpha} \wedge \psi_{+}\,^{\beta}\right) +
\left(A\, m'\, n' + B\, o'\, p'\right) \left(\psi_{+}\,^{\alpha} \wedge \psi_{-}\,^{\beta}\right)\right. \\[0.3cm]
&&\ds\;\;\qquad +\left.\left(A\, n'\, m' + B\ p'\, o'\right) \left(\psi_{-}\,^{\alpha} \wedge \psi_{+}\,^{\beta}\right) \right.\\[0.3cm]
&&\ds\;\;\qquad\qquad +\left.
\left(A\, n'\, n' + B\ p'\, p'\right) \left(\psi_{-}\,^{\alpha} \wedge \psi_{-}\,^{\beta}\right)\right\}\, \Gamma^{a}_{\alpha \beta}
\\[0.35cm]
&&\ds- \frac{i}{16 \cdot (5!)^2} \left\{
 \left(A\, i'\, i' + B\, k'\, k' \right) \left( \rho_{+}\thinspace^{a_{1} \dots a_{5}} \wedge \rho_{+}\thinspace^{b_{1} \dots b_{5}}\right) \right.
 \\[0.3cm]
&&\ds\;\;\qquad\left. +
\left(A\, i'\, j' + B\ k'\, l' \right) \left( \rho_{+}\thinspace^{a_{1} \dots a_{5}} \wedge \rho_{-}\thinspace^{b_{1} \dots b_{5}}\right)
 \right.\\[0.3cm]
&&\ds\;\;\qquad\qquad +\left. \left(A\, j'\, i' + B\, l'\, k' \right) \left( \rho_{-}\thinspace^{a_{1} \dots a_{5}} \wedge \rho_{+}\thinspace^{b_{1} \dots b_{5}}\right) \right.
 \\[0.3cm]
 &&\ds\;\;\qquad\qquad\qquad + \left.
\left(A\, j'\, j' + B\, l'\, l' \right) \left( \rho_{-}\thinspace^{a_{1} \dots a_{5}} \wedge \rho_{-}\thinspace^{b_{1} \dots b_{5}}\right)\right\}\, \epsilon^{a}\thinspace_{a_{1} \dots a_{5}b_{1} \dots b_{5}}\ .
\end{array}\]
To obtain the explicit form of $d{\rho_{-}^{a}}$
is enough to change in the previous expression of $d{\rho_{+}^{a}}$ the $A$ and $B$ for $C$ and $D$, respectively,
i.e.
 \[
d{\rho_{+}^{a}}\;\; \longleftrightarrow\;\; d{\rho_{-}^{a}}\qquad \Longleftrightarrow\qquad
  A\leftrightarrow C,\; B\leftrightarrow D\ .
\]\medskip

{\bf 2).- For two-index tensors}
\[\begin{array}{lll}
d{\rho_{+}^{ab}} &=&\ds - \frac{1}{8}
\left\{ \left( E\, a'\, a' + F\, c'\, c' \right) \left( \rho_{+}\,^{a} \wedge \rho_{+}\,^{b}\right) + \left( E\, a'\, b' + F\, c'\ d' \right) \left( \rho_{+}\,^{a} \wedge \rho_{-}\,^{b}\right) \right.
\\[0.3cm]
&&\qquad\ds\left. +\left(E\, b'\, a' + F\, d'\, c' \right) \left( \rho_{-}\,^{a} \wedge \rho_{+}\,^{b}\right) +
\left( E\, b'\, b' + F\, d'\, d' \right) \left( \rho_{-}\,^{a} \wedge \rho_{-}\,^{b}\right) \right\}
 \\[0.35cm]
&&\ds- \frac{1}{8}
\left\{ \left( E\, e'\, e' + F\, g'\, g'\right) \left( \rho_{+}\,^{ac} \wedge \rho_{+}\,_{c}\,^{b}\right) +
\left(E\, e'\, f' + F\, g'\, h'\right) \left( \rho_{+}\,^{ac} \wedge \rho_{-}\,_{c}\,^{b}\right)\right.\\[0.3cm]
&&\qquad\ds\left. +
 \left( E\, f'\, e' + F\, h'\ g'\right) \left( \rho_{-}\,^{ac} \wedge \rho_{+}\,_{c}\,^{b}\right) +
\left( E\ f'\, f' + F\, h'\ h'\right) \left( \rho_{-}\,^{ac} \wedge \rho_{-}\,_{c}\,^{b}\right) \right\}
\\[0.35cm]
&&\ds- \frac{1}{2}\
\left\{ \left( E\, m'\, m' + F\, o'\, o'\right) \left( \psi_{+}\,^{\alpha} \wedge \psi_{+}\,^{\beta}\right) +
\left( E\, m'\, n' + F\, o'\, p'\right) \left( \psi_{+}\,^{\alpha} \wedge \psi_{-}\,^{\beta}\right) \right.
\\[0.3cm]
&&\qquad\ds\left.+
 \left( E\, n'\, m' + F\, p'\, o'\right) \left( \psi_{-}\,^{\alpha} \wedge \psi_{+}\,^{\beta}\right) +
\left(E\, n'\, n' + F\, p'\, p'\right) \left( \psi_{-}\,^{\alpha} \wedge \psi_{-}\,^{\beta}\right) \right\}\,
\Gamma^{ab}_{\alpha \beta}\\[0.35cm]
&&\ds- \frac{1}{8\cdot 4!}  \left\{ \left(E\, i'\, i' + F\, k'\, k' \right) \left( \rho_{+}\,^{a}\,_{b_{1} \dots b_{4}} \wedge
\rho_{+}\,^{b_{1} \dots b_{4}\,b}\right)\right.\\[0.3cm]
&&\qquad\ds\left. +
\left(E\, i'\, j' + F\, k'\, l' \right) \left( \rho_{+}\,^{a}\,_{b_{1} \dots b_{4}} \wedge \rho_{-}\,^{b_{1} \dots b_{4}\,b}\right)\right.
\\[0.3cm]
&&\qquad\qquad\ds\left.+ \left(E\, j'\, i' + F\, l'\ k' \right) \left( \rho_{-}\,^{a}\,_{b_{1} \dots b_{4}} \wedge \rho_{+}\,^{b_{1} \dots b_{4}\,b}\right)\right.\\[0.3cm]
&&\qquad\qquad\qquad\ds\left.  +
\left(E\,j'\, j' + F\, l'\, l' \right) \left( \rho_{-}\,^{a}\,_{b_{1} \dots b_{4}} \wedge \rho_{-}\,^{b_{1} \dots b_{4}\,b}\right)
\right\}\ .
\end{array}\]

Like in the previous case we have also that
 \[
d{\rho_{+}^{ab}}\;\; \longleftrightarrow\;\; d{\rho_{-}^{ab}}\qquad \Longleftrightarrow\qquad
  E\leftrightarrow G,\;\; F\leftrightarrow H\ .
\]\medskip

{\bf 3).- For five-index tensors}
\[
\begin{array}{lll}
d\rho_{+}^{a_{1} \dots a_{5}}
&=&\ds
- \frac{5}{8} \left\{
\left(I\,e'\, i' + J\, g'\, k' \right) \left( {\rho_{+}}\,^{[a_{1}}\,_{b} \wedge \rho_{+}\,^{b\, a_{2} \dots a_{5}]}\right) \right.\\[0.3cm]
&&\qquad\ds\left. +
\left(I\, e'\, j' + J\, g'\, l' \right) \left( \rho_{+}\,^{[a_{1}}\,_{b} \wedge {\rho_{-}}^{b\,a_{2} \dots a_{5}]}\right) \right.
\\[0.30cm]
&&\ds \qquad\qquad +\left.\left(I\, f'\, i' + J\, h'\ k' \right) \left({\rho_{-}}\,^{[a_{1}}\,_{b} \wedge \rho_{+}\,^{b\,a_{2} \dots a_{5}]}\right) \right.\\[0.3cm]
&&\qquad\quad\qquad\ds\left. + \left(I\, f'\, j' + J\ h'\, l' \right) \left({\rho_{-}}\,^{[a_{1}}\,_{b} \wedge \rho_{-}\,^{b\, a_{2} \dots a_{5}]}\right)\right\}
\\[0.35cm]
&&\ds
- \frac{1}{2}\left\{
\left(I\ m'\ m' + J\ o'\ o'\right) \left(\psi_{+}\,^{\alpha} \wedge \psi_{+}\,^{\beta}\right) +
\left(I\ m'\ n' + J\ o'\ p'\right) \left(\psi_{+}\,^{\alpha} \wedge \psi_{-}\,^{\beta}\right)\right.
\\[0.30cm]
&&\ds\qquad +\left.
\left(I\, n'\, m' + J\, p'\, o'\right) \left(\psi_{-}\,^{\alpha} \wedge \psi_{+}\,^{\beta}\right) \right.\\[0.3cm]
&&\qquad\qquad\ds\left. +
\left(I\, n'\, n' + J\, p'\, p'\right) \left(\psi_{-}\,^{\alpha} \wedge \psi_{-}\,^{\beta}\right)
\right\}\, \Gamma^{a_{1} \dots a_{5}}_{\alpha \beta}
\\[0.35cm]
&&\ds
- \frac{i}{8\cdot 5!}
\left\{
\left(I\ a'\ i' + J\ c'\ k' \right) \left(\rho_{+}\,^{a} \wedge \rho_{+}\,^{b_{1} \dots b_{5}}\right)\right.
\\[0.30cm]
&&\ds \qquad +\left.
\left(I\, a'\, j' + J\, c'\, l' \right) \left(\rho_{+}\,^{a} \wedge \rho_{-}\,^{b_{1} \dots b_{5}}\right)\right.
\\[0.30cm]
&&\ds \qquad +\left. +
\left(I\, b'\, i' + J\, d'\, k' \right) \left(\rho_{-}\,^{a} \wedge \rho_{+}\,^{b_{1} \dots b_{5}}\right)\right.
\\[0.30cm]
&&\ds\left.\qquad\qquad +\left(I\ b'\, j' + J\,
d'\, l' \right) \left(\rho_{-}\,^{a} \wedge \rho_{-}\,^{b_{1} \dots b_{5}}\right)
\right\}\,
\epsilon_{a}\,_{b_{1} \dots b_{5}}\,^{a_{1} \dots a_{5}}
\\[0.35cm]
&&\ds
- \frac{i}{2\cdot (4!)^2}
\left\{
\left( I\, i'\, i' + J\ k'\, k' \right) \left( {\rho_{+}}\,_{b_{1} \dots b_{5}} \wedge {\rho_{+}}\,^{b_{5}b_{4}c_{1}c_{2}c_{3}}\right)
\right. \\[0.3cm]
&&\qquad\ds+\left.
\left( I\, i'\, j' + J\, k'\, l' \right) \left( {\rho_{+}}_{\, b_{1} \dots b_{5}} \wedge {\rho_{-}}^{\, b_{5}b_{4}c_{1}c_{2}c_{3}}\right) \right.
\\[0.30cm]
&&\ds \qquad\qquad\left.+
\left( I\, j'\, i' + J\, l'\, k' \right) \left({\rho_{-}}\,_{b_{1} \dots b_{5}} \wedge {\rho_{+}}^{\, b_{5}b_{4}c_{1}c_{2}c_{3}}\right)
 \right.
\\[0.30cm]
&&\qquad\qquad\qquad\ds +\left.
\left( I\, j'\, j' + J\, l'\, l' \right) \left({\rho_{-}}_{\, b_{1} \dots b_{5}} \wedge {\rho_{-}}^{\,b_{5}b_{4}c_{1}c_{2}c_{3}}\right)
\right\}
\epsilon^{a_{1} \dots a_{5}b_{1}b_{2}b_{3}}_{\,c_{1}c_{2}c_{3}}\, .
\end{array}\]
Also
 \[
d{\rho_{+}}^{a_{1} \dots a_{5}}\;\; \longleftrightarrow\;\; d{{\rho_{-}}^{a_{1} \dots a_{5}}}\qquad
\Longleftrightarrow\qquad
  I\leftrightarrow K,\; \; J\leftrightarrow L
\]\medskip

{\bf 4).- For spinors}
\[\begin{array}{lll}
d\psi_{+}^{\alpha}& =&\ds
 \frac{1}{16} \left\{
\left(M\ m'\ a' + N\ o'\ c'\right) \left(\psi_{+}\,^{\beta} \wedge \rho_{+}\,^{a}\right) +
\left(M\ m'\ b' + N\ o'\ d'\right) \left(\psi_{+}\,^{\beta} \wedge \rho_{-}\,^{a}\right) \right.\\[0.3cm]
&&\ds\qquad \left. +\left(M\ n'\ a' + N\ p'\ c'\right) \left(\psi_{-}\,^{\beta} \wedge \rho_{+}\,^{a}\right)\right.\\[0.3cm]
&&\ds\qquad\qquad \left. +
\left(M\ n'\ b' + N\ p'\ d'\right) \left(\psi_{-}\,^{\beta} \wedge \rho_{-}\,^{a}\right) \right\}\, {\Gamma_{a}}\,_{\beta}\,^{\alpha}
\\[0.35cm]
&&\ds -\frac{1}{32} \left\{
\left(M\, m'\, e' + N\, o'\ g'\right) \left(\psi_{+}\,^{\beta} \wedge \rho_{+}\,^{ab}\right) +
\left(M\ m'\ f' + N\ o'\ h'\right) \left(\psi_{+}\,^{\beta} \wedge \rho_{-}\,^{ab}\right)\right.\\[0.3cm]
&&\ds\qquad+\left. \left(M\ n'\ e' + N\ p'\ g'\right) \left(\psi_{-}\,^{\beta} \wedge \rho_{+}\,^{ab}\right) \right.\\[0.3cm]
&&\ds\qquad\qquad \left.+
\left(M\ n'\ f' + N\ p'\ h'\right) \left(\psi_{-}\,^{\beta} \wedge \rho_{-}\,^{ab}\right)
\right\}\, {\Gamma_{ab}}\,_{\beta}\,^{\alpha}\\[0.35cm]
&&\ds+\frac{1}{16\cdot 5!}
 \left\{
\left(M\ m'\ i' + N\ o'\ k'\right) \left(\psi_{+}\,^{\beta} \wedge \rho_{+}\,^{a_{1} \dots a_{5}}\right)\right. \\[0.3cm]
&&\ds\qquad +\left.\left(M\ m'\ j' + N\ o'\ l'\right) \left(\psi_{+}\,^{\beta} \wedge \rho_{-}\,^{a_{1} \dots a_{5}}\right)
\right.\\[0.3cm]
&&\ds\qquad\qquad \left.+
\left(M\ n'\ i' + N\ p'\ k'\right) \left(\psi_{-}\,^{\beta} \wedge \rho_{+}\,^{a_{1} \dots a_{5}}\right)\right.\\[0.3cm]
&&\ds\qquad\qquad\qquad \left.+\left(M\ n'\ j' + N\ p'\ l'\right) \left(\psi_{-}\,^{\beta} \wedge \rho_{-}\,^{a_{1} \dots a_{5}}\right)
\right\} \, {\Gamma_{a_{1} \dots a_{5}}}\,_{\beta}\,^{\alpha} \ .
\end{array}\]
Finally
 \[
d{\psi_{+}^{\alpha}}\;\; \longleftrightarrow\;\; d{\psi_{-}^{\alpha}}\qquad \Longleftrightarrow\qquad
  M\leftrightarrow O,\; N\leftrightarrow P .
\]


\end{document}